\documentclass[10pt, conference, letterpaper]{IEEEtran}

\usepackage{amsmath,amsfonts}
\usepackage{algorithm,algorithmic}
\usepackage{array}
\usepackage{textcomp}
\usepackage{stfloats}
\usepackage{xurl}
\usepackage{hyperref}
\usepackage{verbatim}
\usepackage{graphicx}
\usepackage{cite}
\usepackage{xcolor}
\usepackage{subfigure}

\usepackage[normalem]{ulem}
\useunder{\uline}{\ul}{}
\usepackage{lscape}

\begin{document}

\title{One-Way Delay Model for 5G Low Latency, High Data Rate Communications}

\author{\IEEEauthorblockN{Hamid Hassani, Sonia M. Heemstra de Groot, Ignas G. Niemegeers, Kishor C. Joshi, George Exarchakos}
	\IEEEauthorblockA{Eindhoven University of Technology, The Netherlands}
        {\tt \{h.hassani,sheemstradegroot,i.g.m.m.niemegeers,k.c.joshi,g.exarchakos\}@tue.nl}
}

\maketitle

\begin{abstract}
Achieving low delay and high data rate communication in 5G requires preventing persistent queue backlogs at path bottlenecks. % at base stations (BSs). 
A key indicator of queue buildup is an increase in one-way delay (OWD) of packets. 
However, accurately measuring OWD is challenging due to synchronization issues between the sender and receiver. 
Moreover, using OWD to infer the onset of queueing %at the path bottleneck 
can be inefficient. For instance, congestion control mechanisms that rely on latency as a congestion signal may result in cellular network underutilization, see, e.g., \cite{Xie22}. 
Thus, to develop more efficient rate control approaches, it is crucial not only to estimate OWD accurately but also to perform a more fine-grained analysis that provides detailed insights into the queue status. % at the path bottleneck. 
This paper presents a framework for modeling OWD in 5G edge networks, where the wireless link is typically the path bottleneck. %, by modeling it through measurable delay components. 
We demonstrate that OWD is dominated by queueing delay (QD), particularly at high data rates. 
We analyze QD by decomposing it into two main components: aggregation delay and channel access time. 
% While queueing delay can be measured precisely at the BS, accessing this information in practice is often challenging, if not impossible. 
Aggregation delay can be accurately measured at the user equipment (UE), and channel access time can be estimated based on network conditions. 
Our key contribution shows that QD, while challenging to measure directly at the base station (BS), can be closely approximated by combining these components. 
We validate this model through NS3 simulations under varying network loads and single- and multi-UE scenarios. %, enabling practical OWD and QD estimation.
This framework provides a foundation for developing advanced mechanisms for congestion control and bottleneck detection in 5G networks.

\end{abstract}

\section{Introduction} \label{sec:intro}
Next-generation applications such as eXtended reality (XR) come with strict requirements for %communication in terms of 
data rate, latency, and reliability, %see, e.g., Table \ref{table:req}, 
which demand a well-designed and well-configured network \cite{Morin22}. 
For instance, \cite{Saad20} states that to provide a full immersive experience, the data rate should range from $100$Gbps to $1$Tbps, latency should be less than $1$ms, and reliability should achieve a level of $5$ to $7$ nines.

% For instance, \cite{Saad20} states that to provide a full immersive experience, the following requirements should be satisfied:
% \begin{table}[hbt]
% \centering
% \caption{QoS Requirements of next-generation XR applications \cite{Saad20}}
% \label{table:req}
% \begin{tabular}{|p{2cm}|p{3.5cm}|}
% \hline
% Data Rate       & $100$Gbps - $1$Tbps       \\ \hline
% Latency         & $< 1$ms                   \\ \hline
% Reliability     & $99.999\%$ - $99.99999\%$ \\ \hline
% \end{tabular}
% \end{table}

These requirements cannot be entirely met by 5G or 802.11ax (Wi-Fi 6) networks \cite{Saad20,Akyildiz22,Reshef22,She23,Deng20}, and many network capabilities will need to grow exponentially during the decade ahead to unleash the full potential of data-intensive technologies \cite{Ericsson,Adeia23,SK23}. 
% \HH{For example, to support advanced XR applications, \cite{Saad20} proposes a new service class for 6G called mobile broadband reliable low-latency communications (MBRLLC), as a generalization of URLLC and eMBB services of 5G.}
% \HH{Achieving optimum performance is an overall goal realized when all layers of the network protocol stack work together \cite{Tanenbaum21}. }
While much attention has been and is being focused on physical and link layer standards, primarily by 3GPP and IEEE 802.11 working groups, it is essential to have well-designed transport layer protocols to avoid congestion collapse and satisfy the stringent requirements of next-generation applications \cite{Hassani21}. For example, \cite{Xie22} has experimentally compared the end-to-end performance of multiple congestion control algorithms using the Pantheon platform \cite{Yan18} in a commercial LTE network (AT\&T), and showed the variability of performance in terms of data rate and one-way delay (OWD) for the different congestion control mechanisms might be significant. 
% This measurement, along with many similar measurements conducted using 3GPP or IEEE 802.11 access technologies, see, e.g., \cite{Xie22,Hassani21,PBE20,ExLL18,Kim14}, provide additional evidence regarding the insufficiency of relying exclusively on physical and link layer solutions.

% The requirement for major upgrades to current transport protocols is also reflected in initiatives such as L4S \cite{L4SNokia}, Google QUIC \cite{QUIC}, Google BBR \cite{BBR}, and the OpenFastPath Alliance \cite{OFA}. Furthermore, ETSI has recently set up the Industry Specification Group on Next Generation Protocols (NGP ISG) to study the next-generation protocols for 5G \cite{NGP}.

According to the applied congestion signal\footnote{An indicator used by congestion control mechanisms to detect congestion within the network.}, the congestion control mechanisms at the transport layer are categorized into two main groups: loss-based and delay-based approaches. For loss-based solutions, e.g., TCP Cubic~\cite{Ha08} and TCP BIC~\cite{Xu04}, packet loss signals congestion. In contrast, delay-based solutions such as TCP BBR~\cite{BBR} and TCP Vegas \cite{vegas}  % TCP NewReno \cite{RFC6582} 
depend on delay measurements, e.g., round-trip time (RTT), to infer the queue buildup at the path bottleneck. %\HH{Another notable approach for congestion control in cellular networks is NG-Scope~\cite{Xie22}, which employs several passive sniffers located in different sites to measure cell-wide resource utilization and gauge available capacity. }
There are alternative approaches, such as using network-wide sniffing~\cite{PBE20,Xie22,Wan24a} or explicit congestion notification (ECN) protocols like L4S~\cite{L4SNokia}, but these solutions are often impractical due to deployment and upgrade difficulties. 

Using packet loss as a congestion signal is not suitable for next-generation applications that would require both high data rate and low latency, as it triggers action, i.e., adjusting the send rate, only after queue overflow has occurred, resulting in significant queueing delay (QD) \cite{Tanenbaum21}. 
% \HH{Furthermore, the transport layer solutions rely on packet loss as the congestion signal cannot differentiate if the loss is due to poor channel quality or queue overflow.} 
On the other hand, for delay-based approaches, the use of RTT %(/min RTT) 
to estimate the onset of queueing is also known to be inaccurate when there is queueing in the reverse path, i.e., asymmetric forward and reverse paths \cite{Chefrour21}. 
Measurements of OWD might be used to infer the onset of queueing and adjust the send rate, but measuring OWD is known to be challenging due to, e.g., clock offset and skew between server and client \cite{Hassani20}, as is inference of queueing from OWD.

To address the challenges mentioned above, we introduce a framework for modeling packet delay in 5G edge architectures, %, as illustrated in Fig. \ref{fig:SysMod} (a)
where the path bottleneck is the radio link control (RLC) buffer at the base station (BS). 
Note that queueing delay is a major component of end-to-end delay \cite{L4SNokia,Irazabal22,Irazabal24}, however, it is typically measurable only at the BS, which is often challenging, if not impossible, in practice. % mainly in high data rates
Our main contribution is a framework for measuring queueing delay with acceptable precision. We demonstrate that the queueing delay at 5G BSs consists of two primary components: aggregation delay and channel access time. We then propose approaches to measure (/estimate) these components. We validate our model through NS3 simulations across various scenarios. 

The remainder of this paper is structured as follows: 
In Section \ref{sec:relatedwork}, we discuss related work. We recap some essential preliminaries in Section \ref{sec:prelim}. 
Next, we present the system architecture in Section \ref{sec:system}, followed by an analysis of latency in Section \ref{sec:delay}. 
In Section \ref{sec:performance}, we present the NS3 simulation results. 
We conclude the paper in Section \ref{sec:conclusion}. 
% \HH{A list of abbreviations used in this paper is provided in Appendix \ref{sec:abb}.}

\section{Related Work} \label{sec:relatedwork}

Previous work on OWD analysis is primarily divided into data-driven and model-driven approaches. 

As a prevalent data-driven approach, machine learning (ML) methods leverage historical data to identify and predict delay patterns. 
For instance, in \cite{Rao24}, domain adversarial neural networks were applied to estimate OWD in 5G networks by analyzing data collected at BSs. 
\cite{Rao22} investigated RTT violation prediction in 5G networks with multiple ML algorithms to showcase their utility in delay prediction. 
While these approaches offer valuable insights, they are highly sensitive to the quality of training data and the availability of large training datasets, which are often challenging to obtain from commercial networks. 
Additionally, the high computational complexity of ML models can lead to increased delay \cite{Lopez23}, particularly when deploying large models on devices with limited processing capabilities, which may not always be suitable for low latency applications. 
Furthermore, their performance may vary due to variations in user equipment (UE) capabilities, which are often not uniform across commercial mobile devices \cite{Rao24}.

Model-driven approaches, based on queueing theory and stochastic network calculus, use probability distributions and stochastic processes to set a theoretical basis for latency analysis. 
For example, \cite{Perales23} proposed a statistical model to predict OWD in V2X applications within 5G networks, while \cite{Kountouris18} used the effective capacity framework to derive an upper bound on delay violation probability, providing insights into delay dynamics. 
\cite{Karnam25} expanded the proposed model in \cite{Kountouris18} by developing a reliability model for beyond 5G networks, incorporating retransmissions, multiple delay components, and interactions across network layers. 
However, such models may have limited accuracy in capturing complex and dynamic systems \cite{Lopez23}, posing challenges for real-time adaptation.

% \HH{The distinctive feature of our work, compared to existing solutions, is that it introduces a straightforward yet effective model that leverages measurements at the UE to accurately approximate OWD in 5G. This approach avoids the main limitations typically associated with data- and model-driven approaches, such as computational complexity and scalability challenges. }
% Our model offers valuable insights for achieving low delay, high data rate communication in 5G networks.

\section{Preliminaries} \label{sec:prelim}

% \subsection{Physical Resource Block}
% PRBs are the minimum resource units that can be allocated to UEs. In the frequency domain, a PRB consists of $12$ subcarriers \cite{ETSI_38.211}. However, in the time domain, the minimum duration of one PRB in 5G can be as short as one OFDM symbol, while its maximum can be up to $14$ OFDM symbols (/one slot\footnote{Note that, with extended CP, each slot contains $12$ OFDM symbols.}, similar to LTE). Furthermore, the duration of a PRB can vary based on, e.g., numerology. A key motivation behind introducing such flexibility is to enable services, such as time-critical applications, that demand low latency communication.

\subsection{Transport Block (TB)}

\begin{figure}[tbp]
\centering
\includegraphics[width=0.95\linewidth]{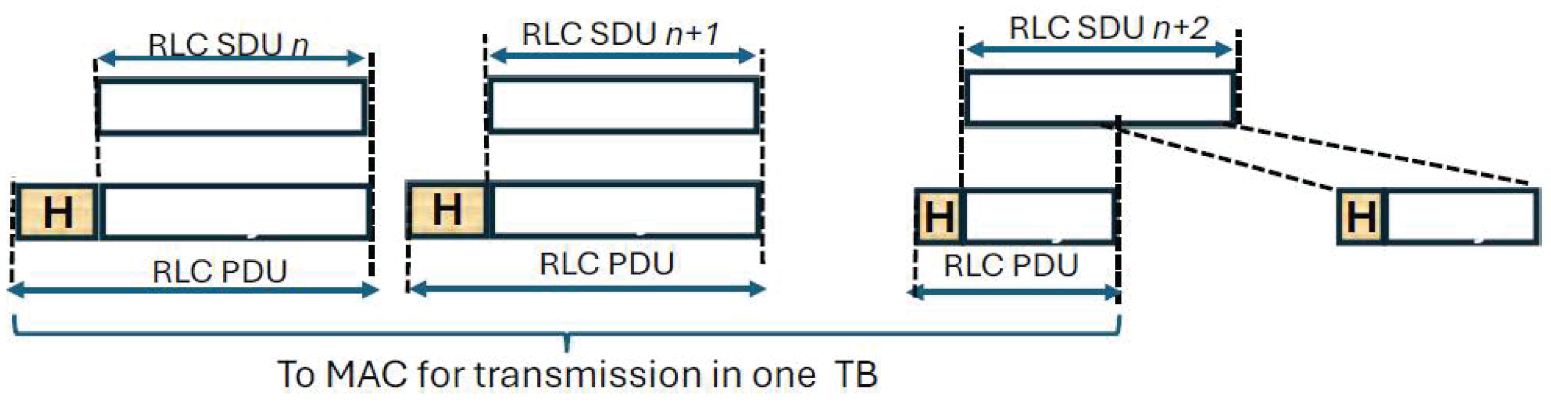}
\caption{RLC segmentation in 5G, adapted from \cite{Dahlman23}.}
\label{fig:seg}
\end{figure}

TBs are used to refer to MAC PDUs\footnote{The data entity from/to a higher protocol layer is termed service data unit (SDU), while the corresponding entity to/from a lower protocol layer is called a protocol data unit (PDU) \cite{Dahlman23}.} (/PHY SDUs) in LTE/5G networks. 
% Note that the MAC scheduler follows a per-UE queueing policy. Therefore, 
After receiving the allocated transport block size (TBS) information for each UE from the MAC scheduler, the RLC layer selects a specific number of packets and attaches the header to each packet to construct RLC PDUs. These packets are then delivered to the MAC layer for transmission within one TB\footnote{Note that up to two transport blocks per UE can be used in the case of spatial multiplexing with more than four layers \cite{Dahlman23}.}. 
To maximize bandwidth efficiency and fill up the TB effectively, the last RLC PDU may contain a segment of an SDU \cite{Dahlman23}, see Fig. \ref{fig:seg}. 
Note that the TBS is determined by the number of allocated physical resource blocks (PRBs) and the modulation and coding scheme (MCS) \cite{Flavien22}.

\subsection{RLC Queue} \label{sec:rlc}
There are at least three levels of queueing within 5G networks: i) the RLC sublayer, ii) the SDAP sublayer, and (iii) the UPF network function buffers \cite{Irazabal22}. 
However, since the wireless link is typically significantly slower than the wired domain, the path bottleneck typically lies at the entity that holds the last buffer in the wired domain, i.e., the RLC buffer \cite{Irazabal22,Irazabal24}. 
Note that for each connected UE, there are one or more PDU sessions, each with one or more QoS flows and data radio bearers (DRBs) in the user plane, responsible for carrying the user data \cite{Dahlman23}. 5G networks support up to $8$ DRBs per UE, with each DRB having its dedicated RLC buffer in both the uplink and downlink \cite{Irazabal22}.

\section{System Architecture} \label{sec:system}
\begin{figure}[tbp]
\centering
\hspace{-25pt}
\subfigure[System model]{
\includegraphics[width=0.62\columnwidth]{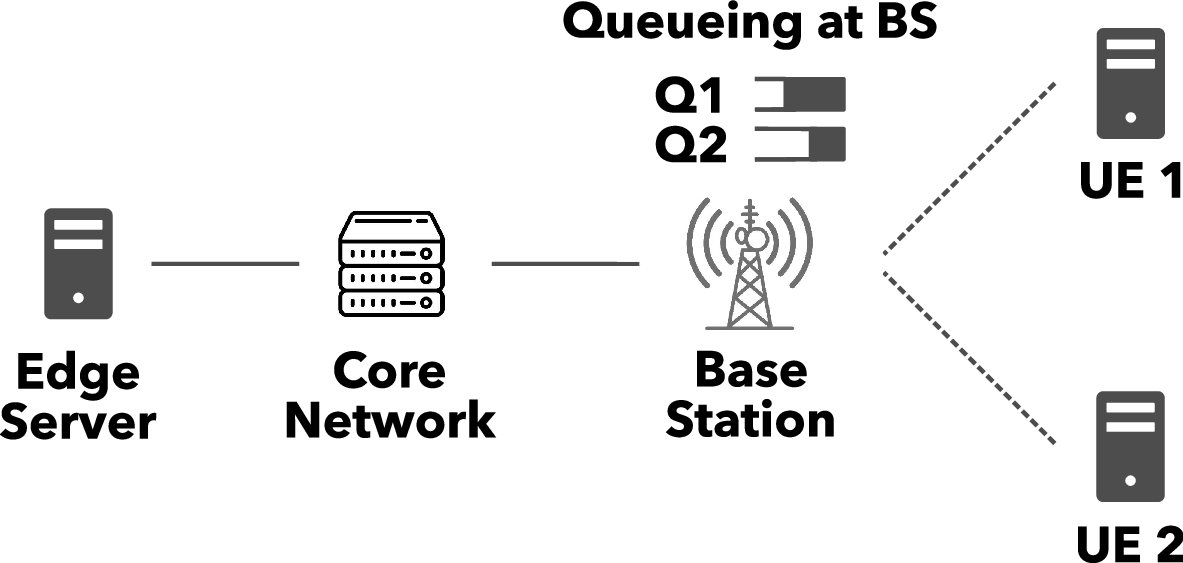}
}
\hspace{-5pt}
\subfigure[User-plane protocol stack]{
\includegraphics[width=0.32\columnwidth]{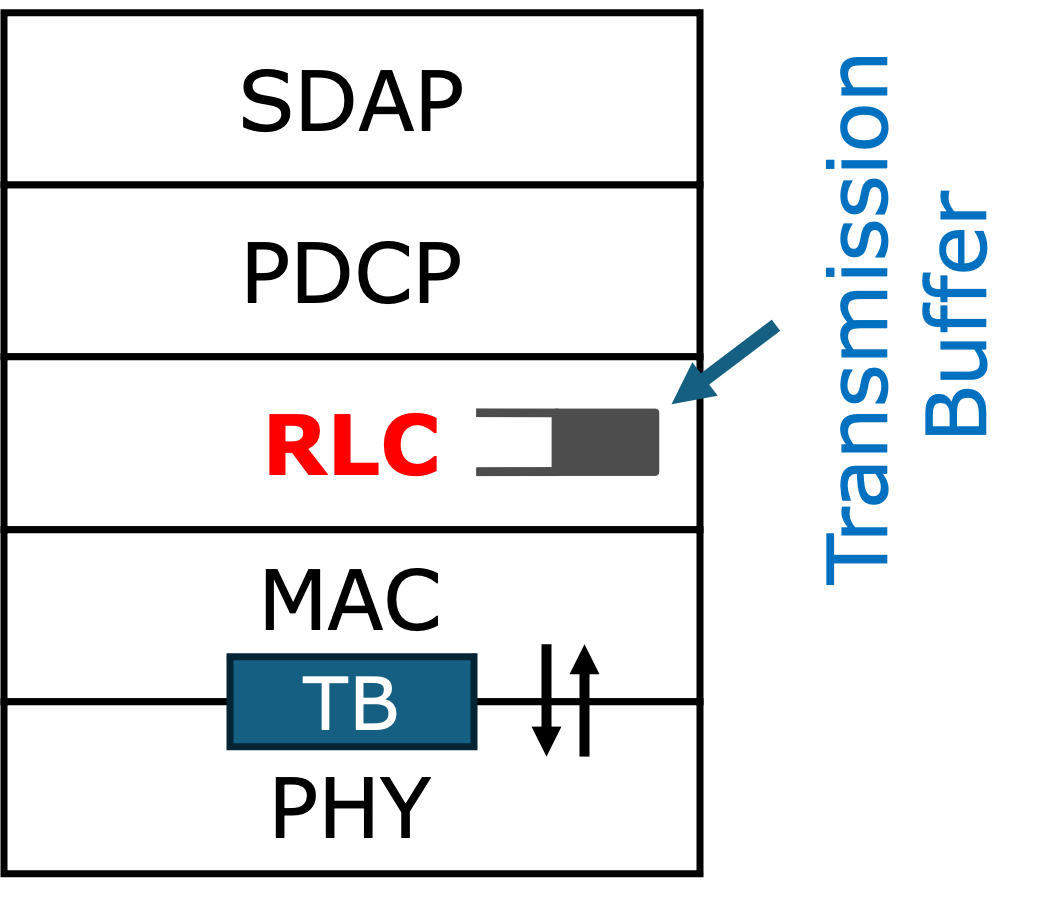}
}
\hspace{-25pt}
\caption{(a) Edge transport architecture with a bottleneck in the wireless link, i.e., at the RLC layer of the BS, as indicated on the schematic. (b) Abstract resource allocation schematic at the BS.}
\label{fig:SysMod}
\end{figure}

In this paper, we consider the edge architecture of the type illustrated in Fig. \ref{fig:SysMod} (a), where the BS serves as the bottleneck in an end-to-end path from the edge server (ES) to UEs. This is due to the rate mismatch between the wired backhaul and the air interface. As depicted in Fig. \ref{fig:SysMod}, the BS adheres to the per-UE queueing policy, meaning that each UE has its dedicated queue at the BS. % This motivates us to focus on client-side solutions.

% Our interest is in achieving high data rate and low latency in 5G edge networks by effectively managing end-to-end connections via an innovative transport layer approach deployed in the user space. 
% The key concept involves preventing the buildup of a persistent queue backlog at the BS in the RLC buffer to keep the end-to-end delay low. 
% Note that the only effective way to manage congestion is for the transport protocols to decrease the rate at which packets are sent into the network \cite{Tanenbaum21}. 

\subsection{Assumptions} \label{sec:assump}
\subsubsection{System Model}
%The network under consideration is a public 5G network where not all traffic flows (/UEs) are under the administration of the edge server. 
The network under consideration is a 5G network, where the edge server is located close to the end users, i.e., at the network edge. %, and has a centralized overview of network flows under its administration. 
The path bottleneck is located in the wireless link. 
% \HH{Furthermore, the backhaul network and ES are assumed to be reliable without any packet losses.} 
% In addition, the UEs are considered semi-static, operating without requiring handovers.

\subsubsection{Data Flow} 
% We use the UDP transport protocol to deploy our approach as an application that runs on both the ES and UEs. 
% The downlink traffic type is considered identical, indicating that all traffic flows toward a UE are carried by a single DRB, thus sharing the same buffer at the RLC layer. 
The downlink traffic can be periodic or bursty but, without loss of generality, is considered identical, indicating that all traffic flows toward a UE are carried by a single DRB, thus sharing the same buffer at the RLC layer. 
Note that the proposed model applies to a single RLC queue, and since traffic flows carried by different DRBs are stored in dedicated buffers, our approach remains applicable to non-identical traffic flows as well. 
% We assume %packets are generated periodically, and 
% that packets within a TB experience consistent transmission time from the ES to the BS, traveling along the same path. 
We assume an ideal backhaul, and that packets within a TB experience consistent transmission time from the ES to the BS, traveling along the same path. 
% No packet re-ordering occurs during the transmission from the ES to the BS. 
% \HH{Furthermore, all data packets are of the same size and} 
Furthermore, all data packets can fit within one maximum transmission unit (MTU) without requiring segmentation. 
We also assume that the RLC layer operates in unacknowledged mode (UM), which does not support packet retransmission and is used for real-time services. 

\section{Latency Analysis} \label{sec:delay}

%To design an effective low latency rate control policy, 
We focus on the one-way delay of packets transmitted from the ES to UEs within the 5G edge network presented in Fig. \ref{fig:SysMod} (a). 
% Recall that measuring RTT is not an accurate approach to infer the onset of queueing at the BS, see Section \ref{sec:intro}. 
The one-way delay includes several random components such as processing time, queueing delay, transmission delay, and alignment delay \cite{Liberg19}. 
However, %as discussed in Section \ref{sec:rlc},
the most significant (and often overlooked) source of delay is typically the queueing delay \cite{L4SNokia} at the RLC layer of the BS \cite{Irazabal22,Flavien22}. 
% Our analysis, therefore, excludes other sources of delay. 
Note that, for ease of presentation, we will refer to this source of delay as queueing delay unless stated otherwise.

% Processing time (across various network layers and nodes), Queueing delay at the RLC layer, UPF, SDAP laye, etc, Transmission delay, Propagation delay, Retransmission delay

The queueing delay of a packet is the duration it spends waiting at the RLC queue of the BS until it is delivered to the MAC layer for transmission over the air interface. 
This period is influenced by various factors, including the underlying access technology and the number of earlier-arriving packets in the queue awaiting transmission over the wireless link. 
Furthermore, the queueing delay is significantly affected by the MAC scheduling policy, the queue occupancy of other users, buffer sizes, and the channel quality of connected UEs to the BS.
% Recall that the BS follows the per-UE queueing policy. 

% https://gaia.cs.umass.edu/kurose_ross/interactive/qdelay.php
Suppose all RLC buffers at the BS are empty, and no other packet is being transmitted. In that case, the queueing delay of any incoming packets will be nearly zero or, at worst, bounded by one scheduling interval (/slot duration). %, see Section \ref{sec:prelim}. 
However, the queueing delay will be prolonged in the presence of substantial (cross-)traffic and a large backlog of packets waiting for transmission.

% \begin{figure}[htp]
% \centering
% \includegraphics[width=0.5\linewidth]{TB.png}
% \caption{Abstract resource allocation schematic at a 5G BS.} % Note that the SDAP layer is only for 5G networks.}
% \label{fig:TB}
% \end{figure}

Fig. \ref{fig:SysMod} (b) depicts the transmission buffer at the RLC layer as well as the cross-layer communication between the MAC and PHY layers to encapsulate queued packets into TBs for transmission. 
% \HH{Hamid: since the RLC layer is also involved in the process, it may be better to add similar arrows for the RLC-MAC interactions as well.}). 
Note that the scheduler is implemented at the MAC layer. 
In each scheduling interval, the scheduler assesses the buffer occupancy of connected UEs and subsequently initiates requests to the PHY layer to allocate resources (/PRBs) to the identified queues. 
% Note that the scheduling interval is fixed at $1$ms in LTE and variable in 5G depending on the numerology, see Section \ref{sec:intro}. However, in 5G, it remains unchanged throughout the system's runtime. 
The scheduling interval is fixed at $1$ms in LTE and variable in 5G depending on, e.g., numerology and scheduling policy \cite{ETSI_38.912}. 
% however, it remains unchanged throughout the system's runtime.}

\subsection{Modeling Delay}
The BS adheres to the per-UE queueing policy, meaning that each UE has its dedicated queue at the BS \cite{Zaki15,Brunello21,Irazabal22}. % Therefore, our focus is on modeling the one-way delay in a single UE scenario. 
We proceed by defining the necessary terms required to analyze and model the queueing delay in 5G edge networks, see Table \ref{table:terms}. We assume that, at the MAC layer, packets $\{i, i+1, ..., i+n\}$ are assembled into TB $b$ to be delivered to the PHY layer for transmission over the air interface. We define the aggregation level as the number of packets, i.e., MAC SDUs (/RLC PDUs), that are grouped within one TB. For instance, the aggregation level of TB $b$ is $n+1$. Note that a lower index indicates an earlier arrival time for the corresponding packet. 
% \HH{This assumes no re-ordering of packets occurs from the ES to the BS, see Section \ref{sec:system}.}
{Note that the aggregation level discussed in this paper differs from the PHY layer aggregation level, which is used to adjust the code rate of the PDCCH channel, see, e.g., \cite{Dahlman23}.}

With access to the MAC layer information, the UE can accurately measure the aggregation level, as all packets aggregated in a TB share the same MAC timestamps, i.e., they arrive at the MAC layer of the UE at the same time. 
This can be achieved, e.g., by using open-source UEs such as OpenAirInterface \cite{OAI} and srsRAN \cite{srs} or software tools like Mobileinsight \cite{Li16}, which run on commercial off-the-shelf mobile handsets. 

\begin{table}[tbp]
\centering
\caption{Definitions of key notations used in modeling delay.}
\label{table:terms}
\renewcommand{\arraystretch}{1.3}
\resizebox{1.0\columnwidth}{!}{
\begin{tabular}{|l|l|}
\hline
$ T_{\text{snd,ES},j} $   & Send time of packet $j$ at the edge server \\ \hline
$ T_{\text{arr,RLC},j} $  & Arrival time of packet $j$ at the RLC queue (BS) \\ \hline
$ T_{\text{dep,RLC},j} $  & Departure time of packet $j$ from the RLC queue (BS) \\ \hline
$ \text{AD}_j $           & Aggregation delay of packet $j$ \\ \hline
$ T_{\text{access},b} $   & Channel access time of TB $b$ \\ \hline
$ \text{QD}_j $           & Queueing delay of packet $j$ \\ \hline
% $ T_{\text{arr,MAC},j} $  & Arrival time of packet $j$ at the MAC layer (BS) \HH{seems unused} \\ \hline
% $ T_{\text{rec,UE},j} $   & Receive time of packet $j$ at the UE \HH{seems unused} \\ \hline
\end{tabular}}
\end{table}

% Fig. \ref{fig:timeline} provides a visual representation of the terms defined in Table \ref{table:terms}. 
% As depicted in Fig. \ref{fig:timeline}, TB $b$ comprises $n+1$ packets indexed from $i$ to $i+n$. These packets are simultaneously delivered to the PHY layer for transmission, therefore, all have the same departure time from the RLC queue:
% \begin{equation} \label{eq:Tdep}
%     T_{\text{dep, RLC}, i} = T_{\text{dep, RLC}, i+1} = ... = T_{\text{dep, RLC}, i+n} = T_{\text{dep, RLC}, b}
% \end{equation}
% where $T_{\text{dep, RLC}, b}$ denotes the departure time of TB $b$ from the RLC layer of the BS, see Fig. \ref{fig:timeline}.

% \HH{Is it better to use $T_{\text{dep, RLC}, i, b}$ to denote packet $i$ in TB $b$?}

% A similar concept applies to the arrival time of these aggregated packets at the MAC layer of the UE (/receiver): \HH{seems unnecessary!}
% \begin{equation}
%     T_{\text{rec, MAC}, i} = T_{\text{rec, MAC}, i+1} = ... = T_{\text{rec, MAC}, i+n} = T_{\text{rec, MAC}, b}
% \end{equation}

Fig. \ref{fig:timeline} provides a visual representation of the terms defined in Table \ref{table:terms}. 
As depicted in Fig. \ref{fig:timeline}, we define the queueing delay of packet $j$ that is assembled into TB $b$ %at the BS 
as follows:
% packet $j \in \{i, i+1, ...., i+n\}$ in TB $b$ as follows:
\begin{equation} \label{eq:qd}
\begin{split}
    \text{QD}_j = \text{AD}_j + T_{\text{access},b} %\,\,\, \text{for} \,\,\, j \in \{i, i+1, ..., i+n\}
\end{split}
\end{equation}
where $\text{AD}_j$ denotes the aggregation delay of packet $j$, and $T_{\text{access},b}$ represents the channel access time of TB $b$, as depicted in Fig. \ref{fig:timeline}. 
% Note that the first packet in a TB experiences the largest delay compared to the other packets within that TB.

% Next, we introduce the following terms that are essential for modeling the queueing delay : (i) $\text{AD}_j$, the aggregation delay of packet $j$, and (ii) $T_{\text{access},b}$, the channel access time of TB $b$. 

\subsubsection{\textbf{Aggregation Delay}}
$\text{AD}_j$ represents the time that packet $j \in \{i, i+1, ..., i+n\}$ waits in the RLC queue until the arrival of the last packet in TB $b$. This is defined as follows:
\begin{equation} \label{eq:cd1}
\begin{split}
    \text{AD}_j = T_{\text{arr, RLC}, i+n} - T_{\text{arr, RLC}, j} % \\ \, \, \,\,\, \text{for} \,\,\, j \in \{i, i+1, ..., i+n\}
\end{split}
\end{equation}
As all packets within a TB experience an identical delay from the ES to the BS and no re-ordering occurs in the physical link, see Section \ref{sec:system}, we can rewrite \eqref{eq:cd1} as follows:
\begin{equation} \label{eq:cd2}
\begin{split}
    \text{AD}_j & \approx T_{\text{snd,ES}, i+n} - T_{\text{snd,ES}, j} \\
    & \quad \text{for} \,\,\, j \in \{i, i+1, ..., i+n\}
\end{split}
\end{equation}
Note that the first packet in a TB experiences the largest aggregation delay compared to the other packets within that TB, while 
%Note that 
the aggregation delay of the last packet in a TB is zero. %\HH{, e.g., $\text{AD}_{i+n} = 0$. (remove)} 
% We \HH{also (remove)} assume that the ES adds $T_{\text{snd,ES}, j}$ to the payload of packet $j$ using the system epoch/kernel time. Therefore, the UE can extract this timestamp from the packet payload to measure the queueing delay \HH{aggregation delay precisely}. 
The ES adds $T_{\text{snd,ES}, j}$ in the payload of packet $j$ using the system epoch/kernel time. The UE can then extract this timestamp from the packet payload to measure the aggregation delay. 

\subsubsection{\textbf{Channel Access Time}}
$T_{\text{access},b}$ refers to the duration between the arrival of the last packet in TB $b$ at the RLC queue of the BS and the departure of the corresponding TB from that queue. 
% Note that $T_{\text{access},b}$ is the same for all the packets aggregated into TB $b$, see Fig. \ref{fig:timeline}.
We define $T_{\text{access},b}$ as:
\begin{equation} \label{eq:tacc}
\begin{split}
    T_{\text{access},b} &= T_{\text{dep, RLC}, j} - T_{\text{arr,  RLC}, i+n} \\
    & \quad \text{for} \,\,\, j \in \{i, i+1, ..., i+n\} \\
    % & \{\text{using }\eqref{eq:Tdep}\} \\
    % &= T_{\text{dep, RLC}, b} -  T_{\text{arr,  RLC}, i+n}
\end{split}
\end{equation}

\begin{figure}[tbp]
\centering
\includegraphics[width=0.95\linewidth]{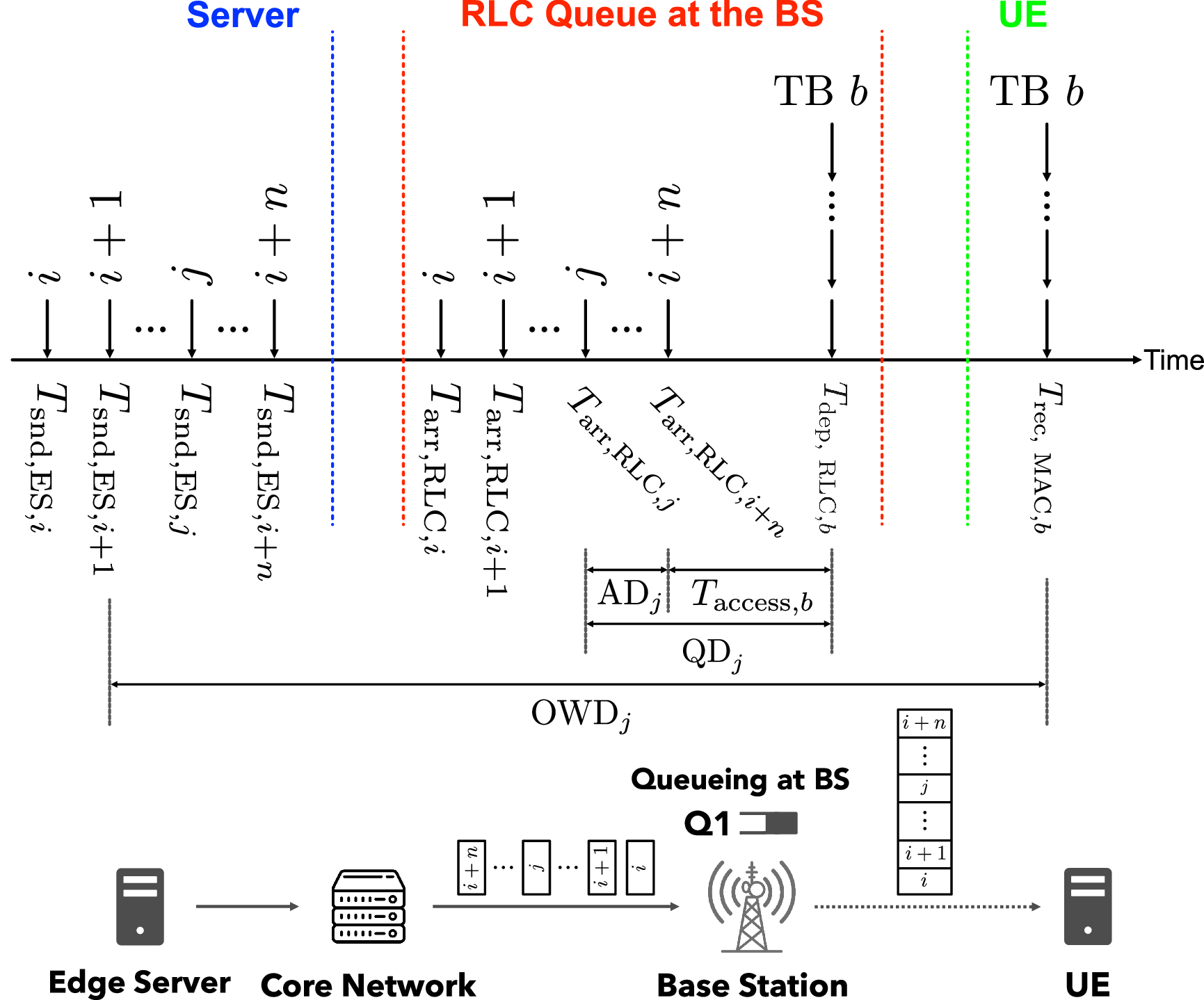}
\caption{Visual representation of terms defined in Table \ref{table:terms}.}
\label{fig:timeline}
\end{figure}

As depicted in Fig. \ref{fig:timeline}, the packets aggregated into TB $b$ are delivered to the PHY layer simultaneously so that they all share the exact departure time from the RLC queue. Therefore, $T_{\text{access},b}$ is identical for all the aggregated packets into TB $b$.

% As depicted in Fig. \ref{fig:timeline}, TB $b$ comprises $n+1$ packets indexed from $i$ to $i+n$. These packets are simultaneously delivered to the PHY layer for transmission, therefore, all have the same departure time from the RLC queue:
% \begin{equation} \label{eq:Tdep}
%     T_{\text{dep, RLC}, i} = T_{\text{dep, RLC}, i+1} = ... = T_{\text{dep, RLC}, i+n} = T_{\text{dep, RLC}, b}
% \end{equation}
% where $T_{\text{dep, RLC}, b}$ denotes the departure time of TB $b$ from the RLC layer of the BS.

% We can rewrite \eqref{eq:tacc} as follows:
% \begin{equation}
%     T_{\text{access},b} = T_{\text{dep, RLC}, b} -  T_{\text{arr,  RLC}, i+n}
% \end{equation}
% where $T_{\text{dep, RLC}, b}$ denotes the departure time of TB $b$ from the RLC layer of the BS.

For the last packet in TB $b$, i.e., packet $i+n$, we can rewrite \eqref{eq:qd} as follows:
\begin{equation} \label{eq:TaccQD}
\begin{split}
    T_{\text{access},b} &= \text{QD}_{i+n} - \text{AD}_{i+n} \\
    & \{ \text{Using } \eqref{eq:cd2} \text{ with }  j = i+n\} \\
    &\approx \text{QD}_{i+n} - (T_{\text{snd,ES}, i+n} - T_{\text{snd,ES}, i+n}) \\
    &= \text{QD}_{i+n}
\end{split}
\end{equation}
  
We proceed by defining $T_{\text{sys}}$ as the time associated with delay factors that are either fixed or bounded according to NR specifications. These factors include processing time at the BS and UE, transmission time interval (TTI), and alignment delay, which can be determined based on the network configurations like numerology and UE capability, see, e.g., \cite{Nagata17,Liberg19}. 
In Section \ref{sec:performance}, we show that $T_{\text{sys}}$ can be comparable to the QD under low network loads. 

Our focus is on edge architectures where the ES and the UPF are collocated with the BS, see, e.g., Fig. 3 in \cite{ETSI_18}. 
With this architecture, the delay between the ES and BS is typically small and negligible. 
Therefore, we assume that QD is a major component of end-to-end delay and approximate the OWD of a packet $j$ as follows:
\begin{equation} \label{eq:owd}
\begin{split}
    \text{OWD}_{j} \approx \text{QD}_{j} + T_{\text{sys}}
    % \text{OWD}_{i+n} \approx T_{\text{access},b} - T_{\text{sys}}
\end{split}
\end{equation}
Note that if the latency between the ES and BS is significant, we can incorporate it into $T_{\text{sys}}$ as it remains constant or bounded under the assumption of an ideal backhaul network, see Section \ref{sec:assump}.

We can then rewrite \eqref{eq:TaccQD} using \eqref{eq:owd} as follows:
\begin{equation} \label{eq:taccess}
\begin{split}
    T_{\text{access},b} \approx \text{OWD}_{i+n} - T_{\text{sys}}
    % \text{OWD}_{i+n} \approx T_{\text{access},b} - T_{\text{sys}}
\end{split}
\end{equation}
where $\text{OWD}_{i+n}$ is the OWD of the last packet in TB $b$. %, i.e., packet $i+n$. %, and $T_{\text{sys}}$ represents the time associated with factors such as processing time at the BS and UE, TTI, and alignment delay \cite{Nagata17,Liberg19}. 

\subsubsection{Estimating QD}
% \HH{Next, by substituting \eqref{eq:cd2} and \eqref{eq:taccess} into \eqref{eq:qd}, we can approximate the QD of packet $j \in \{i, i+1, ..., i+n\}$ assembled into TB $b$ as follows:
We substitute \eqref{eq:cd2} and \eqref{eq:taccess} into \eqref{eq:qd} and approximate the QD of packet $j \in \{i, i+1, ..., i+n\}$ assembled into TB $b$ as follows:
% \begin{equation} \label{eq:qd1}
% \begin{split}
%     \text{QD}_{j} \approx \big(T_{\text{snd,ES}, i+n} - T_{\text{snd,ES}, j}\big) + \big(\text{OWD}_{i+n} - T_{\text{sys}}\big)
%     % \text{OWD}_{i+n} \approx T_{\text{access},b} - T_{\text{sys}}
% \end{split}
% \end{equation}
% }
\begin{equation} \label{eq:qd1}
\begin{split}
    \text{QD}_{j} \approx \underbrace{\big(T_{\text{snd,ES}, i+n} - T_{\text{snd,ES}, j}\big)}_{\text{AD}} + \underbrace{\big(\text{OWD}_{i+n} - T_{\text{sys}}\big)}_{T_{\text{access}}}
\end{split}
\end{equation}
Note that the first packet in a TB experiences the largest queueing delay compared to the other packets within that TB. 

Using QD instead of RTT %(or OWD) 
provides a more accurate indication of congestion at the path bottleneck. Therefore, \eqref{eq:qd1} can be applied to design low-delay congestion control mechanisms that react more effectively to the onset of queueing at the RLC layer of the BS, facilitating low-delay and high data rate communication in 5G networks. 

% \textcolor{red}{(remove: This equation was Eq. (8) in the main submission.) Next, by substituting \eqref{eq:qd} and \eqref{eq:taccess} into \eqref{eq:owd}, we can approximate the OWD of packet $j \in \{i, i+1, ..., i+n\}$ assembled into TB $b$ as follows:
% \begin{equation} \label{eq:owd2}
% \begin{split}
%     \text{OWD}_{j} \approx \text{AD}_{j} + \text{OWD}_{i+n}
%     % \text{OWD}_{i+n} \approx T_{\text{access},b} - T_{\text{sys}}
% \end{split}
% \end{equation}}
% \textcolor{red}{We can use \eqref{eq:owd2} to closely approximate the OWD of each packet using measurements of aggregation delay.}

The aggregation delay and channel access time are influenced by various factors. These include the arrival rate of packets at the RLC queue, % (/packet send rate at the ES)
the MAC scheduling policy at the BS, cross-traffic load, and the channel quality (/MCS) of UEs connected to the BS.

\subsection{Analyzing $T_{\text{access},b}$} \label{sec:Taccess}

% The challenge is to model (/measure) $T_{\text{access},b}$ in the presence of unknown cross-traffic or interference, particularly in public networks. 
While a precise analysis of $T_{\text{access},b}$ can be challenging, knowing which packets are assembled in one TB can assist in conducting a coarse-grained analysis of it. 
Recall that this information can be readily and accurately measured at the UE, having access to MAC layer information. 
% This can be achieved, e.g., by using open-source UEs such as OpenAirInterface \cite{OAI} and srsRAN \cite{srs} or software tools like Mobileinsight \cite{Li16}, which run on commercial off-the-shelf mobile handsets.

Assume that (i) packets $\{i, i+1, ..., i+n\}$ belong to TB $b$, and (ii) at time $T_{\text{arr,RLC},i+n}$, the last packet of TB $b$, i.e., packet $i+n$, arrived at the RLC queue. 
Furthermore, consider that the next scheduling decision occurs at time $t_0$, where $t_0 \ge T_{\text{arr,RLC},i+n}$. 
Recall that when packets of TB $b$ are not completely delivered to the RLC layer, $T_{\text{access},b}$ is zero, and it starts to increase only when the last packet of TB $b$ arrives at the RLC queue. Therefore, $T_{\text{access},b} = 0$ for $t \le T_{\text{arr,RLC},i+n}$. 
% Note that we can conceptually consider cases where the last packet of a TB arrives at the RLC queue at exactly the scheduling time, i.e., $T_{\text{arr,RLC},i+n} = t_0$. This scenario, however, is deemed infeasible in practice, and we exclude it from our analysis.
   
We define \texttt{SchedIntv} as the scheduling interval of the MAC scheduler, i.e., the duration of one slot. %, see Section \ref{sec:prelim}.
For $t \ge T_{\text{arr,RLC},i+n}$, we divide the network operation into two distinct phases/scenarios:
\begin{itemize}
    \item \textbf{Scenario 1:} At the scheduling decision at $t_0$, the MAC scheduler allocates a sufficient number of PRBs to transmit TB $b$. We, therefore, have:
    \begin{equation*}
        0 \le T_{\text{access},b} < \texttt{SchedIntv}
    \end{equation*}
    
     If $T_{\text{access},b}$ remains within this range for a series of consecutive TBs, it implies that the queue is drained in each scheduling (without persistent queue backlog). 
     % This scenario is considered ideal for achieving low delay communication.
    
    \item \textbf{Scenario 2:} At the scheduling decision at time $t_0$, the TB $b$ cannot be granted transmission, so it should wait for the subsequent scheduling periods, which are set at $t_0 + z * \texttt{SchedIntv}$ where $z \in \{1, 2, \ldots \}$. 
    This situation occurs when, e.g., there are already packets in the queue (arrived earlier), so those packets should be transmitted first prior to the TB of interest. For this scenario, we have:
    \begin{equation*}
        T_{\text{access},b} \ge \texttt{SchedIntv}
    \end{equation*}
\end{itemize}

% Note that Scenario 1 represents an optimal network operation for achieving low delay and high data rate communication in 5G networks.

Scenario 1 represents an optimal network operation, where persistent queue buildup is avoided, which is crucial for achieving low delay and high data rate communication in 5G networks.

% \HH{Hamid: I will add a plot to show the different regions!}

% Note that, unlike the aggregation delay, $T_{\text{access}}$ can be significantly higher, often on the order of several tens of milliseconds, depending on numerology and the duration of each slot. It is influenced by, e.g., the design of the MAC scheduler and cross-traffic. 

% \HH{Hamid: I reserve this topic for a dedicated section/paper focused on our multi-path solution. In that context, I intend to explore $T_{\text{access},b}$ in both 802.11 and 5G networks.}

\subsection{MAC Layer Retransmission}
Using \eqref{eq:qd1}, %\eqref{eq:owd2}, 
we can approximate the QD of packets assembled in a TB only if the UE successfully decodes the TB upon the first transmission attempt. 
However, if the TB is erroneous and the UE is unable to decode it, the UE sends a NACK to the BS, requesting subsequent retransmission(s)\footnote{Note that if the UE still cannot decode a TB after $3$ retransmissions, the TB will be discarded \cite{ETSI_38.212}.}. 

We use the new data indicator (NDI), which is included in the HARQ information of the DL-SCH transport channel, to identify whether a TB is a new transmission or a retransmission. 
If the NDI in the received HARQ information is toggled, this indicates a new transmission; otherwise, it specifies a retransmission of the respective TB \cite{ETSI_38.321}. The sequence number for retransmissions can be determined, e.g., by using the redundancy version (RV) sequence number within the HARQ information. 

The interval from the reception of the first transmission to the reception of the final retransmission represents the ``effective'' delay that contributes to the QD, during which the TB remains in the retransmission buffer at the MAC layer of the BS. 
Note that this duration can be measured accurately on the UE side, as it relies on a single clock reference. 
We %\HH{can} 
then add this delay to \eqref{eq:qd1} to estimate the QD of the packets grouped into that TB. 

% The New Data Indicator (NDI), included in the HARQ information of the DL-SCH or UL-SCH transport channels, indicates whether a TB is a new transmission or a retransmission. 
% If the NDI in the received HARQ information is toggled, this indicates a new transmission; otherwise, it specifies a retransmission of the respective TB \cite{ETSI_38.321}. The sequence number for retransmissions can be determined, e.g., by using the Redundancy Version (RV) sequence number within the HARQ information.

\section{Model Evaluation with NS3} \label{sec:performance}

% \subsection{Configuration}
We use the NS3 network simulator with the 5G-LENA module \cite{Patriciello19} to simulate 3GPP NR non-standalone 5G networks. 
To retrieve the data required for assessing the proposed model, we modified the RLC and MAC layers of 5G-LENA, i.e., \texttt{lte-rlc-um.cc} and \texttt{nr-gnb-mac.cc} scripts. 
The RLC layer operates in UM, meaning there are no packet retransmissions at this layer. Note that the 5G-LENA implementation uses a per-UE queueing policy at the BS. 
% For the propagation loss model, we use the \texttt{InH\_OfficeOpen\_Los} model for a channel
We use a channel with line-of-sight characteristics, i.e., \texttt{InH\_OfficeOpen\_Los}, operating over $40$MHz bandwidth in the n78 frequency band. The numerology is set to $0$, i.e., $15$kHz subcarrier spacing. % similar to LTE, unless stated otherwise. 
We use the default packet scheduler of the 5G-LENA module, i.e., \texttt{NrMacSchedulerTdmaRR}, which allocates available resources equally among UEs \cite{5gLENA}. 
The BS transmission power is set to $10$dBm, and the MCS is fixed at $28$. 
We use the UDP transport protocol to validate our model. Note that the MTU in our setup is configured to $1500$B\footnote{The majority of commercial networks configure their MTU sizes within the range of $1428$ to $1500$ bytes \cite{Osseiran11}.}.

% \subsection{Model Validation}

\begin{figure}[tb]
\centering
\hspace{-25pt}
\subfigure[Single UE]{
\includegraphics[width=0.50\columnwidth]{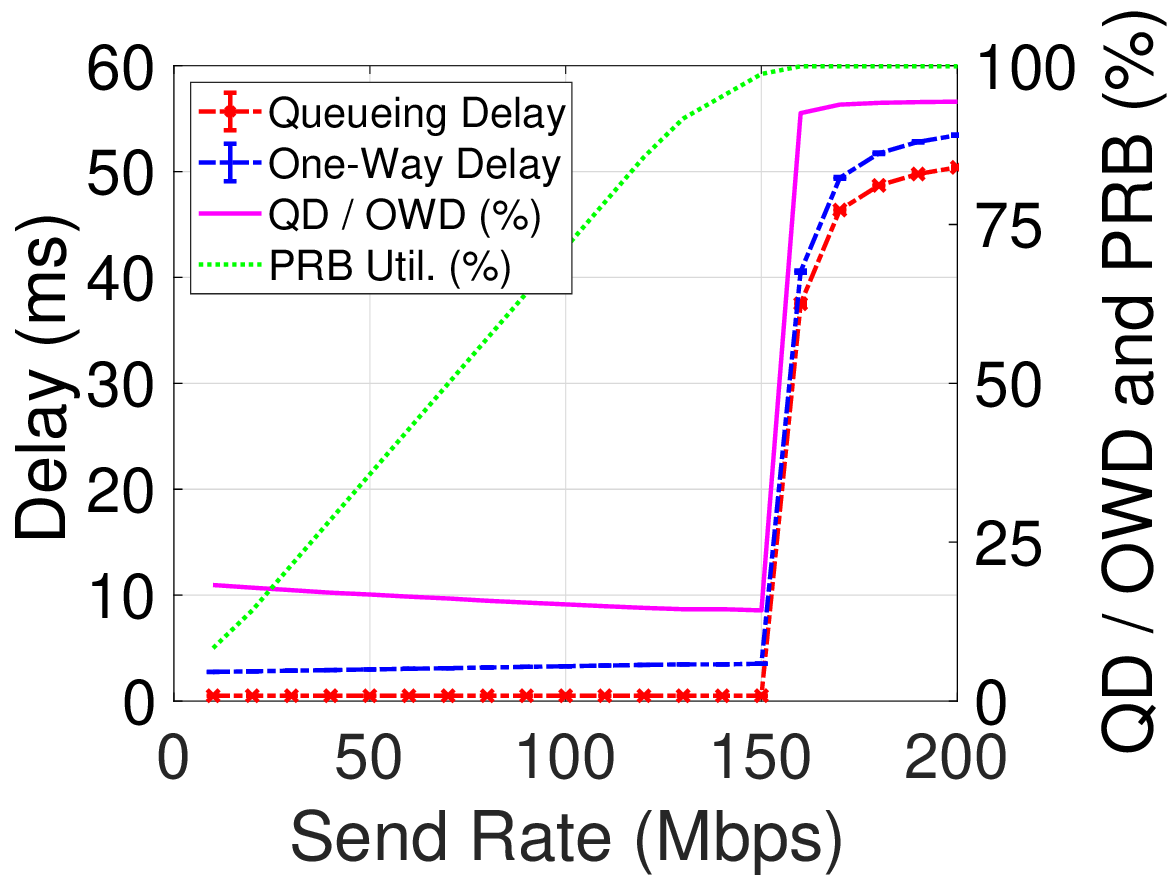}
}
\hspace{-10pt}
\subfigure[Multiple UEs, UDP traffic]{
\includegraphics[width=0.50\columnwidth]{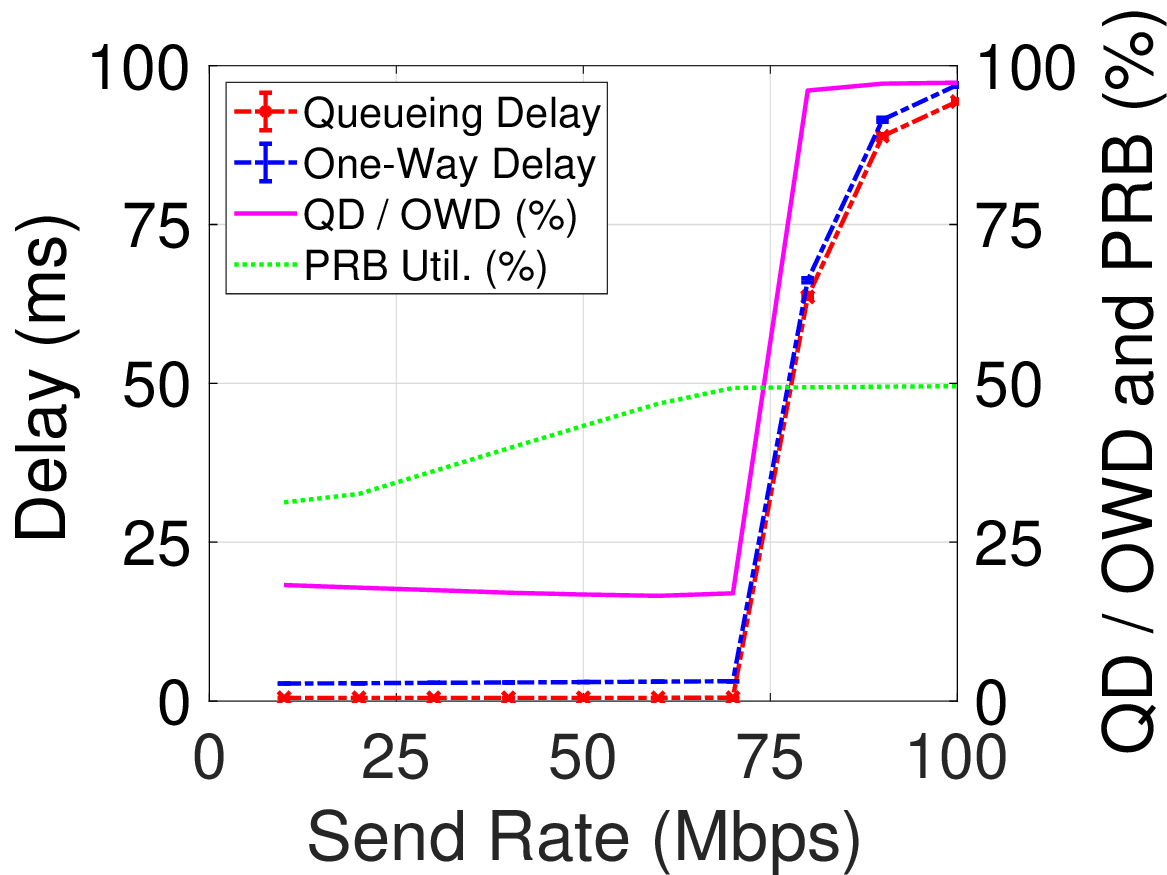}
}
\hspace{-25pt}
\caption{Network performance for (a) a single UE scenario with UDP traffic, (b) $2$ UEs, one using UDP and the other using TCP Cubic. $60$sec of simulation.}
\label{fig:fig3}
\end{figure}

Fig. \ref{fig:fig3} illustrates the network performance for both single- and multi-UE scenarios across varying UDP send rates. 
For the multi-UE case, the results are shown for the UE with UDP traffic, while the TCP flow to the other UE is considered as cross-traffic. 
As we can see in Fig. \ref{fig:fig3} (a) and (b), (i) PRB utilization increases with the data send rate (/network load), and (ii) both OWD and QD also rise gradually as the send rate increases. 
However, when PRB utilization approaches its maximum, there is a sharp increase in delay due to network congestion and queue buildup at the BS. 
Observe in Fig. \ref{fig:fig3} (b) that the PRB utilization of the UE with UDP traffic reaches a peak of $\approx 50\%$ when the network load is high. This is due to the fair scheduling policy applied at the MAC layer. 
As shown in Fig. \ref{fig:fig3}, the proportion of QD within OWD decreases slightly at low network loads, %prior to the sharp increase in high data loads
as the increase in QD is smaller than that in OWD. 
In contrast, near the network saturation point where resource utilization approaches its maximum, the percentage of QD within OWD rises significantly ($> 90\%$) in both scenarios. 
This trend indicates that, at higher network loads, QD becomes the dominant factor contributing to OWD. 
Note that the region close to the network saturation point represents an optimal operating range for high data rate, low latency applications. 
Furthermore, the proportion of QD within OWD is higher in the presence of cross-traffic, see Fig. \ref{fig:fig3} (b), compared to the single-UE scenario depicted in Fig. \ref{fig:fig3} (a). This increase is due to the higher $T_{\text{access}}$ in the multi-UE scenario caused by scheduling latency. % (and identical $T_{\text{sys}}$)

\begin{figure}[tb]
\centering
\hspace{-25pt}
\subfigure[Single UE]{
\includegraphics[width=0.50\columnwidth]{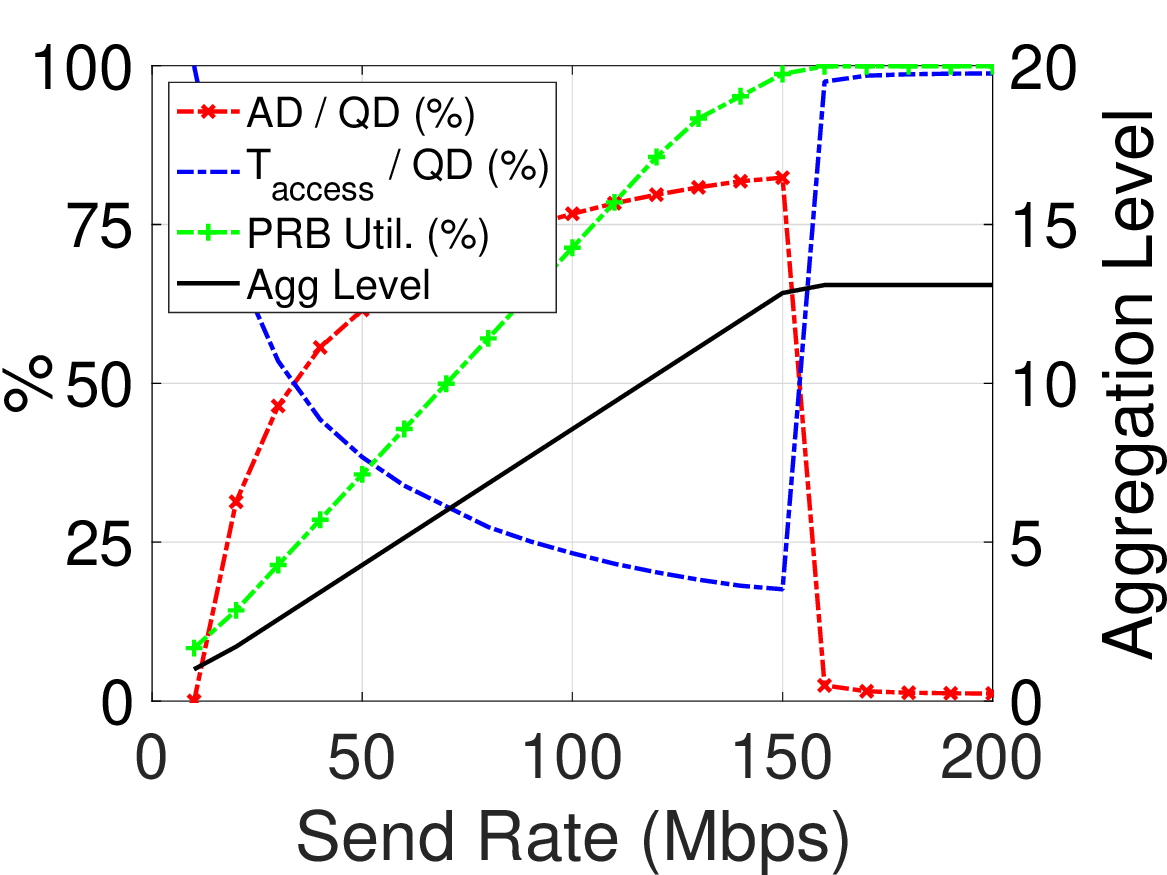}
}
\hspace{-10pt}
\subfigure[Multiple UEs]{
\includegraphics[width=0.50\columnwidth]{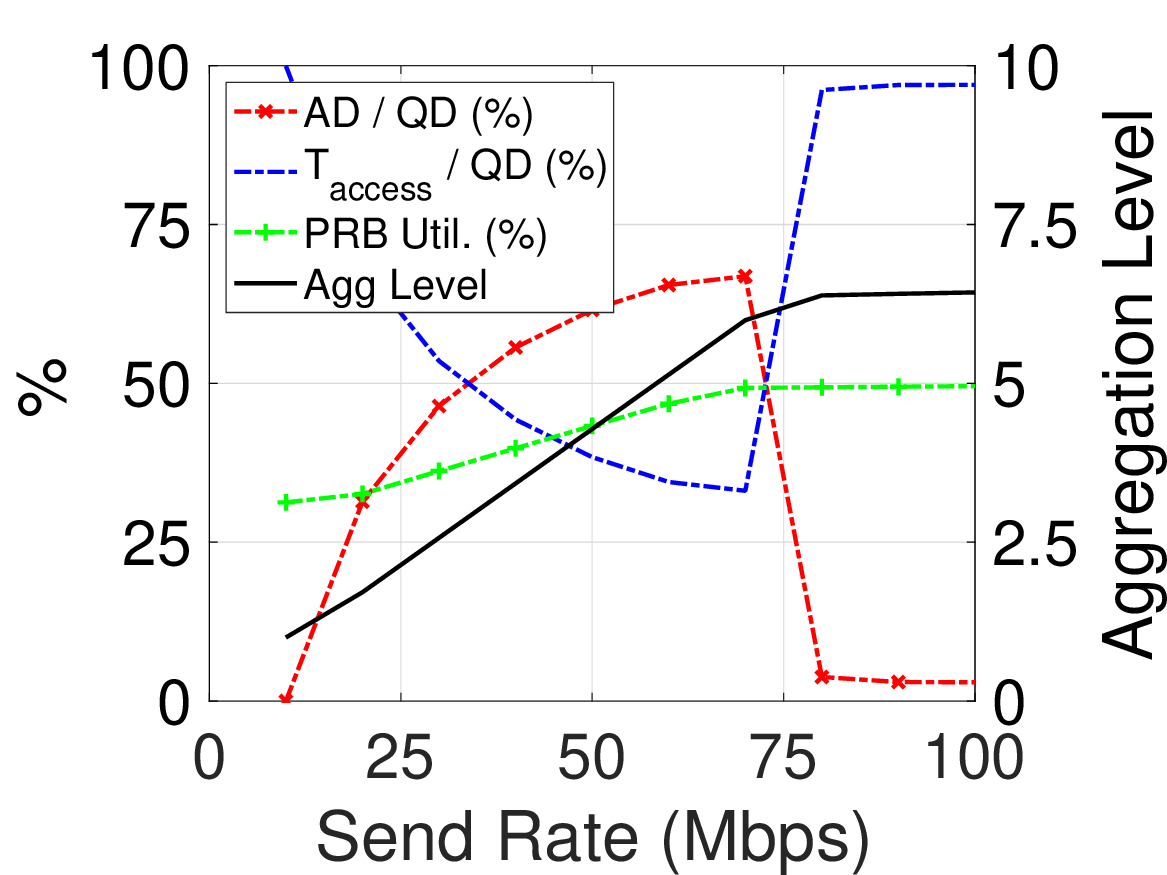}
}
\hspace{-25pt}
\caption{Analysis of queueing delay in (a) a single UE scenario, (b) $2$ UEs, one using UDP and the other using TCP Cubic. $60$sec of simulation.}
\label{fig:fig6}
\end{figure}

Fig. \ref{fig:fig6} demonstrates the contributions of $\text{AD}$ and $T_{\text{access}}$ to QD, represented as $\text{AD} / \text{QD}\,(\%)$ and $T_{\text{access}} / \text{QD}\,(\%)$, respectively, for both single- and multi-UE scenarios over a range of send rates. 
As the send rate increases, distinct trends develop among these metrics. 
As illustrated in Fig. \ref{fig:fig6} (a) and (b), $\text{AD} / \text{QD}$ initially rises due to an increase in the aggregation level, reaching peak levels of $80\%$ for single-UE and $70\%$ for multi-UE scenarios, near maximum PRB utilization. 
In contrast, $T_{\text{access}} / \text{QD}$ declines with increasing the send rate, reaching $20\%$ for single-UE and $30\%$ for multi-UE scenarios, prior to the onset of queueing at the BS. 
However, as PRB utilization increases and the network begins to congest, these trends shift, with $T_{\text{access}}$ emerging as the dominant factor in QD ($> 95\%$). 
These results indicate that to achieve high data rate, low latency communication, it is critical to minimize (/bound) $T_{\text{access}}$ while simultaneously regulating $\text{AD}$ via monitoring the aggregation level of received TBs. 
This corresponds to the network operating region discussed as Scenario 1, see Section \ref{sec:Taccess}. 

\begin{figure}[tbp]
\centering
\hspace{-15pt} %%
\subfigure[Single-UE, $60$Mbps]{
\includegraphics[width=0.50\columnwidth]{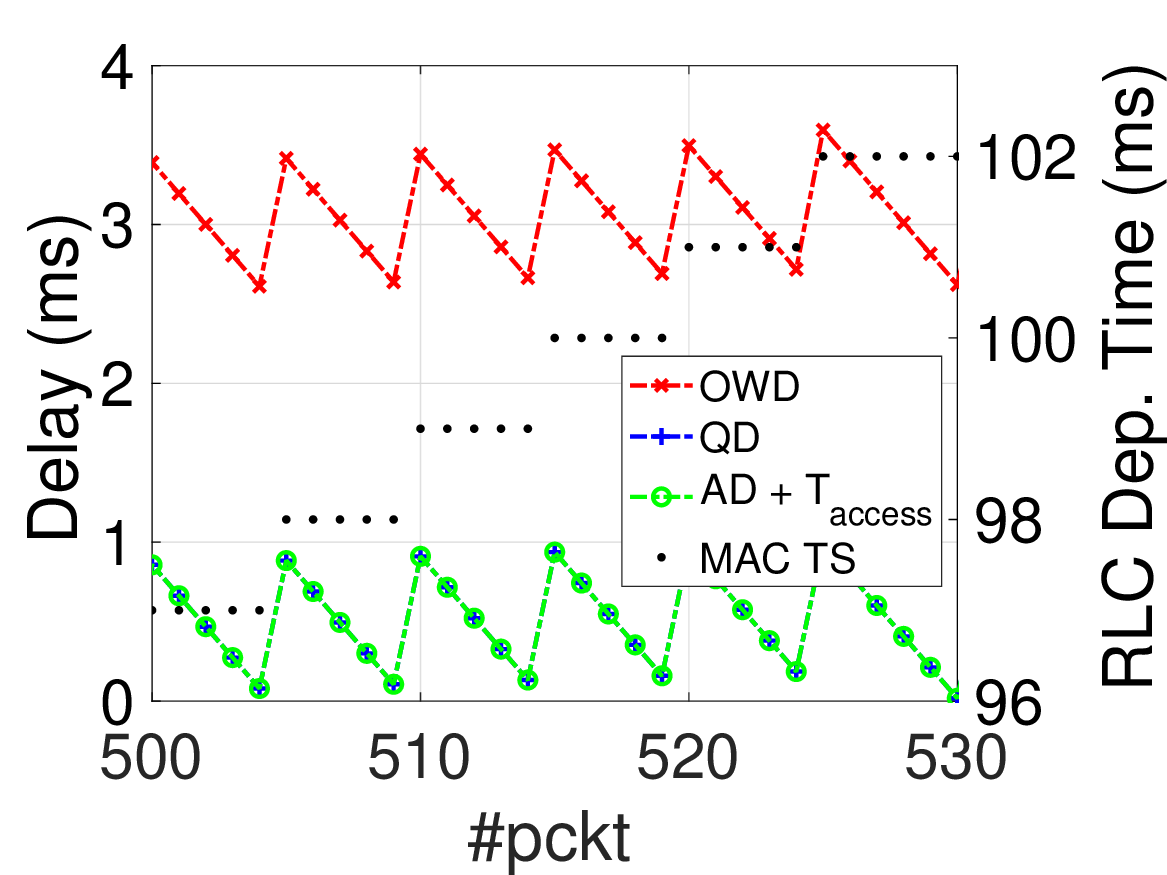}
}
\hspace{-10pt}
\subfigure[Single-UE, $160$Mbps]{
\includegraphics[width=0.50\columnwidth]{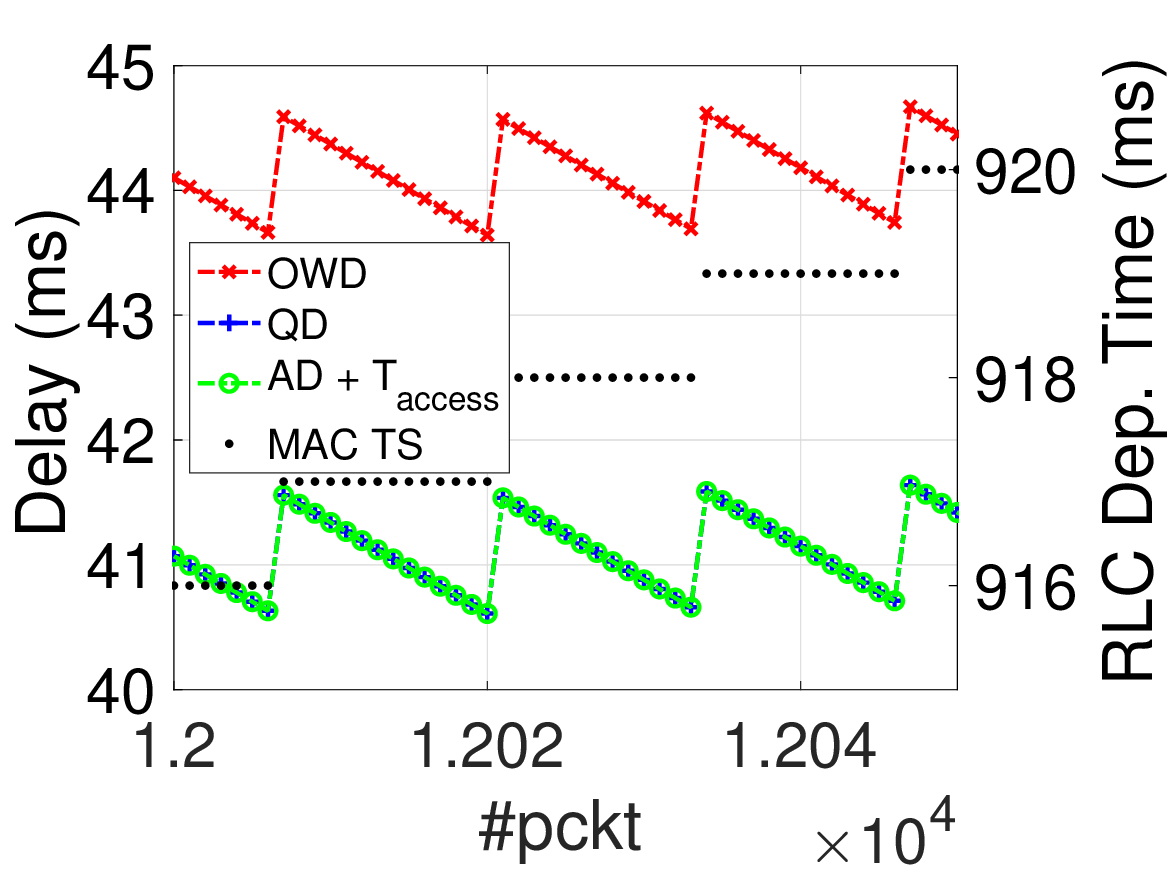}
}
\\
\hspace{-15pt} %%
\subfigure[Multi-UE, $40$Mbps]{
\includegraphics[width=0.50\columnwidth]{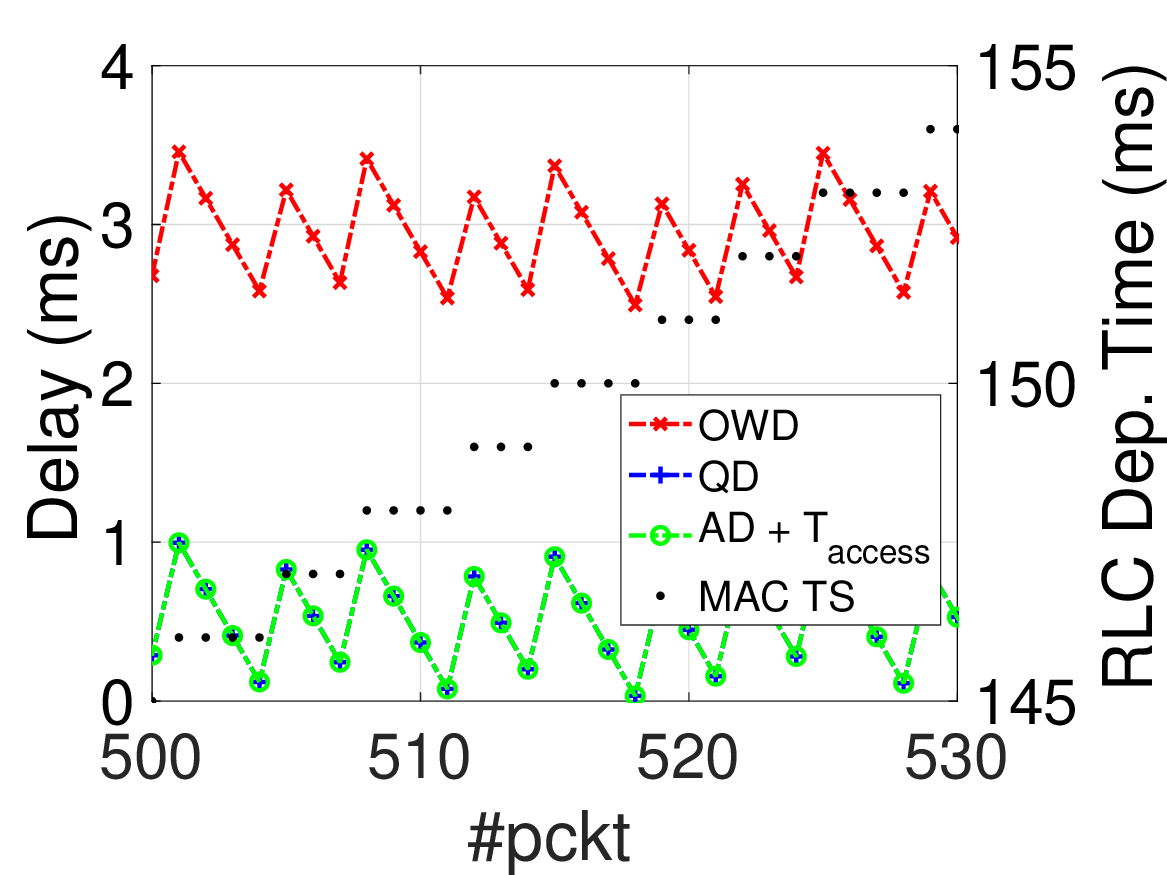}
}
\hspace{-10pt}
\subfigure[Multi-UE, $80$Mbps]{
\includegraphics[width=0.50\columnwidth]{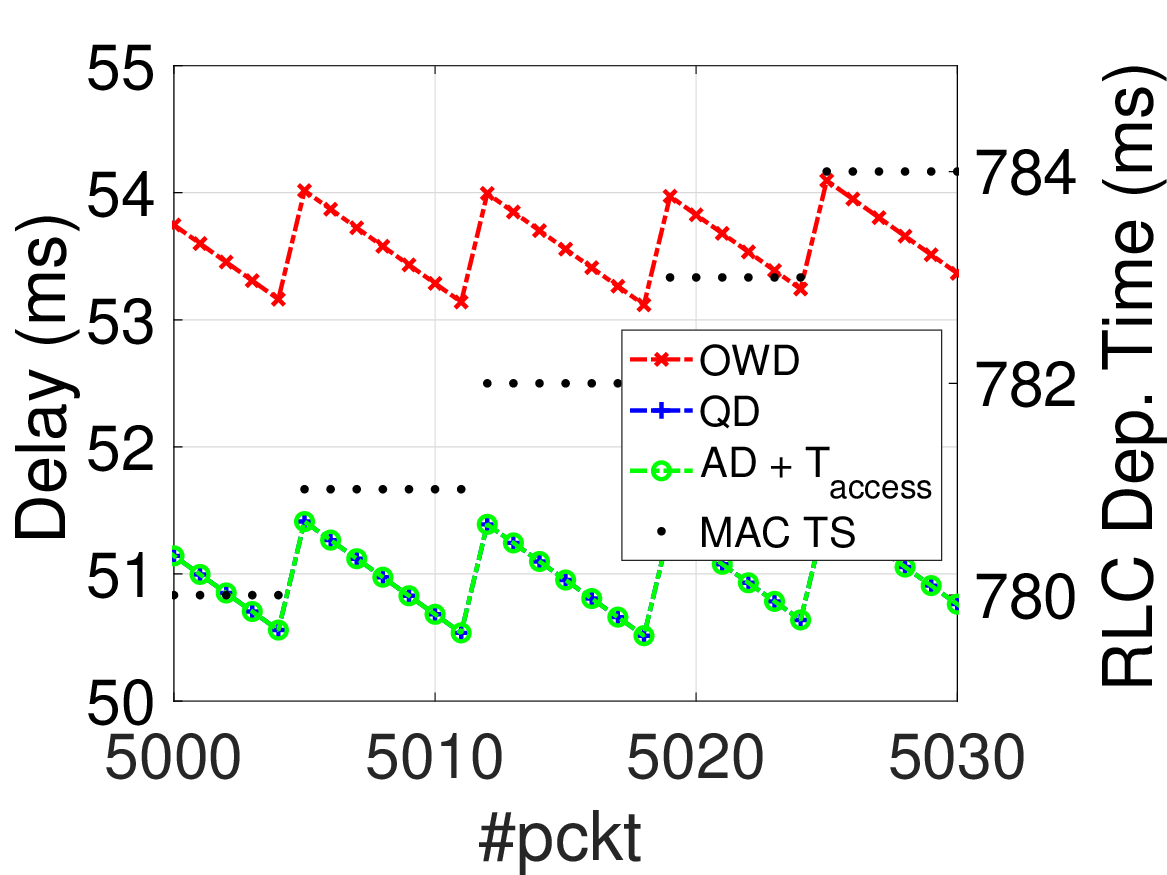}
}
\caption{Model validation in scenarios with (i) a single UE: (a) low network load, (b) high network load; and (ii) $2$ UEs, one using UDP and the other using TCP Cubic: (c) low network load, (d) high network load.}
\label{fig:fig7}
\end{figure}

The validation of the proposed model is illustrated in Fig. \ref{fig:fig7} for both low and high network loads in single- and multi-UE scenarios. 
As shown in Fig. \ref{fig:fig7}, the curve for $\text{AD} + T_{\text{access}}$ matches the QD curve, see \eqref{eq:qd}, across different network loads and scenarios. 
From Fig. \ref{fig:fig7}, we can also observe the following: (i) the first packet in a TB experiences the highest delay, while the last packet encounters the lowest delay within the same TB. Recall that packets within a TB share an identical RLC departure time; (ii) both OWD and QD  exhibit similar patterns. The sawtooth pattern observed in OWD and QD is attributed to the aggregation of multiple packets into a single TB. While these simulations involve periodic packet transmission, a similar pattern is expected for bursty traffic; (iii) there is a persistent gap between OWD and QD, around $2.5$ms, primarily due to constant or bounded sources of latency in 5G networks, such as processing time and TTI \cite{Liberg19}, as denoted by $T_{\text{sys}}$ in Section \ref{sec:delay}. 

% \section{Experimental Evaluation}

\section{Summary and Conclusions} \label{sec:conclusion}

In this paper, we present a novel framework for approximating queueing delay in 5G edge networks. 
We model the queueing delay at the BS as a combination of measurable delay components: aggregation delay and channel access time. 
We then show that the UE can accurately measure the aggregation delay and estimate the channel access time based on network conditions. 
Our findings demonstrate that at %\HH{(remove) high data rates} 
high network loads, queueing delay at the RLC layer of the BS significantly influences OWD. Through NS3 simulations, we validate our model across various network loads and user scenarios, highlighting its potential for practical QD estimation. 
This framework can be used with various traffic types regardless of network load and accounts for retransmissions, providing valuable insights for addressing the critical challenges in achieving low latency and high data rate communication in 5G networks. 
% This framework offers valuable insights for addressing the critical challenges in achieving low latency and high data rate communication in 5G networks.

We plan to experimentally evaluate the accuracy of our model using open-source 5G platforms in different scenarios and traffic types. Furthermore, we aim to investigate the application of this framework in designing an end-to-end rate allocation policy at the transport layer to enable low delay and high data rate communication in 5G edge networks.

\section*{Acknowledgment}
% This work was supported by the Distributed Artificial Intelligent System (DAIS) project under the Key Digital Technologies Joint Undertaking (KDT JU) grant agreement No. 101007273.
This work was supported by the Distributed Artificial Intelligent Systems (DAIS) project under the Key Digital Technologies Joint Undertaking (KDT JU), Grant Agreement No. 101007273. 

% \appendix

% \section{List of Abbreviations} \label{sec:abb}
% \HH{Table \ref{table:abb} provides a list of abbreviations frequently used throughout the paper to assist readers in understanding the technical content.}

% \begin{table}[!h]
% \caption{List of Abbreviations}
% \label{table:abb}
% \centering
% \begin{tabular}{|l|l|}
% \hline
% \textbf{Abbreviation} & \textbf{Definition} \\
% \hline
% AD & Aggregation Delay \\ \hline
% BS & Base Station \\ \hline
% ES & Edge Server \\ \hline
% OWD & One-Way Delay \\ \hline
% PRB & Physical Resource Block \\ \hline
% QD & Queueing Delay \\ \hline
% TB & Transport Block \\ \hline
% UE & User Equipment \\ \hline
% \end{tabular}
% \end{table}

\bibliographystyle{IEEEtran}
\bibliography{references}

@article{Saad20,
    author={Saad, Walid and Bennis, Mehdi and Chen, Mingzhe},
    journal={IEEE Network}, 
    title={{A Vision of 6G Wireless Systems: Applications, Trends, Technologies, and Open Research Problems}}, 
    year={2020},
    volume={34},
    number={3},
    pages={134-142}}

@online{Ericsson,
    title={{Future Network Trends Driving Universal Metaverse Mobility}},
    url={https://www.ericsson.com/en/reports-and-papers/ericsson-technology-review/articles/technology-trends-2022},
    year={2022},
    note={Accessed: May 10, 2026}}

@article{Morin22,
    author={Morin, Diego Gonzalez and Perez, Pablo and Armada, Ana Garcia},
    journal={IEEE Communications Magazine}, 
    title={{Toward the Distributed Implementation of Immersive Augmented Reality Architectures on 5G Networks}}, 
    year={2022},
    volume={60},
    number={2},
    pages={46-52}}

@article{Hassani21,
    author={Hassani, Hamid and Gringoli, Francesco and Leith, Douglas J.},
    journal={Computer Networks},
    pages={107820},
    title={{Quick \& Plenty: Achieving Low Delay \& High Rate in 802.11ac Edge Networks}},
    volume={187},
    year={2021}}

@article{Ha08,
    author={Ha, Sangtae and Rhee, Injong and Xu, Lisong},
    title={{CUBIC: A New TCP-Friendly High-Speed TCP Variant}},
    year={2008},
    volume={42},
    number={5},
    journal={ACM SIGOPS Operating Systems Review},
    pages={64–74},
    month={July}}

@article{Xie22,
    author={Xie, Yaxiong and Jamieson, Kyle},
    title={{NG-Scope: Fine-Grained Telemetry for NextG Cellular Networks}},
    year={2022},
    volume={6},
    number={1},
    journal={Proceedings of the ACM on Measurement and Analysis of Computing Systems},
    month={Feb}}

@article{She23,
    author={She, Changyang and Duong, Trung Q. and Quek, Tony Q.S. and Viswanathan, Harish and Lopez-Perez, David},
    journal={IEEE Wireless Communications}, 
    title={{Guest Editorial: Intelligent Ultra-Reliable and Low-Latency Communications in 6G}}, 
    year={2023},
    volume={30},
    number={2},
    pages={12-13}}

@article{Akyildiz22,
    title={{Wireless Communication Research Challenges for Extended Reality (XR)}},
    author={Akyildiz, Ian F. and Guo, Hongzhi},
    journal={ITU Journal on Future and Evolving Technologies},
    year={2022},
    volume={3}, 
    number={2},
    pages={273-287}}

@article{Reshef22,
    author={Reshef, Ehud and Cordeiro, Carlos},
    journal={IEEE Communications Magazine}, 
    title={{Future Directions for Wi-Fi 8 and Beyond}}, 
    year={2022},
    volume={60},
    number={10},
    pages={50-55}}

@article{BBR,
    author={Cardwell, Neal and Cheng, Yuchung and Gunn, C. Stephen and Hassas Yeganeh, Soheil and Jacobson, Van},
    title={{BBR: Congestion-Based Congestion Control}},
    journal={Communications of the ACM},
    year={2017},
    month={Jan},
    volume={60},
    number={2},
    pages={58–66}}

@book{Tanenbaum21,
    title={{Computer Networks}},
    author={Tanenbaum, Andrew S. and Feamster, Nick and Wetherall, David},
    year={2021},
    edition={6},
    publisher={Pearson}}

@inproceedings{PBE20,
    title={{PBE-CC: Congestion Control via Endpoint-Centric, Physical-Layer Bandwidth Measurements}},
    author={Xie, Yaxiong and Yi, Fan and Jamieson, Kyle},
    booktitle={Proceedings of the ACM SIGCOMM Conference},
    pages={451-464},
    month={Aug},
    year={2020}}

@techreport{Adeia23,
    type={{White Paper}},
    author={{Adeia}},
    title={{Immersive Media: Bringing Science Fiction to Commercial Reality}},
    year={2023},
    month={May},
    institution={Adeia}}

@techreport{SK23,
    type={{White Paper}},
    author={{SK Telecom}},
    title={{5G Lessons Learned, 6G Key Requirements, 6G Network Evolution, and 6G Spectrum}},
    year={2023},
    month={Aug},
    institution={SK Telecom}}

@article{Deng20,
    author={Deng, Cailian and Fang, Xuming and Han, Xiao and Wang, Xianbin and Yan, Li and He, Rong and Long, Yan and Guo, Yuchen},
    journal={IEEE Communications Surveys \& Tutorials}, 
    title={{IEEE 802.11be Wi-Fi 7: New Challenges and Opportunities}}, 
    year={2020},
    volume={22},
    number={4},
    pages={2136-2166}}

@phdthesis{Flavien22,
    author={Ronteix--Jacquet, Flavien},
    title={{Reducing Latency and Jitter in 5G Radio Access Networks}},
    school={{Ecole nationale sup{\'e}rieure Mines-T{\'e}l{\'e}com Atlantique}},
    year={2022}}

@article{Irazabal22,
    author={Irazabal, Mikel and Lopez-Aguilera, Elena and Demirkol, Ilker and Nikaein, Navid},
    journal={IEEE Transactions on Mobile Computing}, 
    title={{Dynamic Buffer Sizing and Pacing as Enablers of 5G Low-Latency Services}}, 
    year={2022},
    volume={21},
    number={3},
    pages={926-939}}

@book{Dahlman23,
    title={{5G/5G-ADVANCED: The New Generation Wireless Access Technology}},
    author={Dahlman, Erik and Parkvall, Stefan and Skold, Johan},
    year={2023},
    edition={3},
    publisher={Elsevier}}

@inproceedings{Yan18,
    author={Yan, Francis Y. and Ma, Jestin and Hill, Greg D. and Raghavan, Deepti and Wahby, Riad S. and Levis, Philip and Winstein, Keith},
    title={{Pantheon: The Training Ground for Internet Congestion-Control Research}},
    year={2018},
    booktitle={Proceedings of the USENIX Annual Technical Conference (USENIX ATC)},
    pages={731–743},
    location={Boston, MA, USA}}

@techreport{ETSI_38.912,
    type={{Technical Specification (TS) 138.912 v17.6.0}},
    author={{ETSI}},
    title={{5G; Study on New Radio (NR) Access Technology}},
    year={2022},
    month={May},
    institution={European Telecommunications Standards Institute (ETSI)}}

@article{Nagata17,
    author={Nagata, Satoshi and Wang, Li Hui and Takeda, Kazuki},
    journal={IEEE Wireless Communications}, 
    title={{Industry Perspectives}}, 
    year={2017},
    volume={24},
    number={3},
    pages={2-4}}

@article{Patriciello19,
    author={Natale Patriciello and Sandra Lagen and Biljana Bojovic and Lorenza Giupponi},
    title={{An E2E Simulator for 5G NR Networks}},
    journal={Simulation Modelling Practice and Theory},
    volume={96},
    pages={101933},
    year={2019}}

@inproceedings{vegas,
    author={Brakmo, Lawrence S. and O'Malley, Sean W. and Peterson, Larry L.},
    title={{TCP Vegas: New Techniques for Congestion Detection and Avoidance}},
    booktitle={Proceedings of the Conference on Communications Architectures, Protocols and Applications (SIGCOMM)},
    pages={24–35},
    year={1994},
    location={London, United Kingdom}}

@inproceedings{Xu04,
    author={Xu, Lisong and Harfoush, K. and Rhee, Injong},
    title={{Binary Increase Congestion Control (BIC) for Fast Long-Distance Networks}}, 
    booktitle={Proceedings of the 23rd Annual Joint Conference of the IEEE Computer and Communications Societies (INFOCOM)}, 
    volume={4},
    pages={2514-2524},
    year={2004},
    location={Hong Kong, China}}

@article{OAI,
    author={Nikaein, Navid and Marina, Mahesh K. and Manickam, Saravana and Dawson, Alex and Knopp, Raymond and Bonnet, Christian},
    title={{OpenAirInterface: A Flexible Platform for 5G Research}},
    journal={ACM SIGCOMM Computer Communication Review},
     pages={33–38},
    volume={44},
    number={5},
    month={Oct},
    year={2014}}

@online{srs,
    title={{5G - srsRAN Project}},
    url={https://www.srslte.com/5g},
    year={2024},
    note={Accessed: May 10, 2026}}

@book{Osseiran11,
    title={{Mobile and Wireless Communications for IMT‐Advanced and Beyond}},
    author={Osseiran, Afif and Monserrat, Jose F and Mohr, Werner},
    year={2011},
    publisher={John Wiley \& Sons}}

@inproceedings{Li16,
    author={Li, Yuanjie and Peng, Chunyi and Yuan, Zengwen and Li, Jiayao and Deng, Haotian and Wang, Tao},
    title={{MobileInsight: Extracting and Analyzing Cellular Network Information on Smartphones}},
    booktitle={Proceedings of the 22nd Annual International Conference on Mobile Computing and Networking (MobiCom)},
    pages={202–215},
    location={New York City, New York},
    year={2016}}

@book{Liberg19,
    title={{Cellular Internet of Things: From Massive Deployments to Critical 5G Applications}},
    author={Liberg, Olof and Sundberg, Marten and Wang, Eric and Bergman, Johan and Sachs, Joachim and Wikstr{\"o}m, Gustav},
    year={2019},
    publisher={Academic Press}}

@article{Chefrour21,
    author={Chefrour, Djalel},
    title={{One-Way Delay Measurement From Traditional Networks to SDN: A Survey}},
    year={2021},
    volume={54},
    number={7},
    journal={ACM Computing Surveys},
    month={July}}

@article{Lopez23,
    author={López, Onel L. A. and Mahmood, Nurul H. and Shehab, Mohammad and Alves, Hirley and Rosabal, Osmel Martínez and Marata, Leatile and Latva-Aho, Matti},
    title={{Statistical Tools and Methodologies for Ultrareliable Low-Latency Communication—A Tutorial}}, 
    journal={Proceedings of the IEEE}, 
    year={2023},
    volume={111},
    number={11},
    pages={1502-1543}}

@inproceedings{Rao24,
    author={Rao, Akhila and Riaz, Hassam and Zavodovski, Aleksandr and Mochaourab, Rami and Berggren, Viktor and Johnsson, Andreas},
    title={{Generalizable One-Way Delay Prediction Models for Heterogeneous UEs in 5G Networks}},
    booktitle={Proceedings of the IEEE Network Operations and Management Symposium (NOMS)},  
    year={2024},
    pages={1-9}}

@inproceedings{Rao22,
    author={Rao, Akhila and T\"{a}rneberg, William and Fitzgerald, Emma and Corneo, Lorenzo and Zavodovski, Aleksandr and Rai, Omkar and Johansson, Sixten and Berggren, Viktor and Riaz, Hassam and Kilinc, Caner and Johnsson, Andreas},
    title={{Prediction and Exposure of Delays From a Base Station Perspective in 5G and Beyond Networks}},
    year={2022},
    booktitle={Proceedings of the ACM SIGCOMM Workshop on 5G and Beyond Network Measurements, Modeling, and Use Cases},
    pages={8–14},
    location={Amsterdam, Netherlands}}

@article{Perales23,
    author={Coll-Perales, Baldomero and Lucas-Estañ, M. Carmen and Shimizu, Takayuki and Gozalvez, Javier and Higuchi, Takamasa and Avedisov, Sergei and Altintas, Onur and Sepulcre, Miguel},
    title={{End-to-End V2X Latency Modeling and Analysis in 5G Networks}}, 
    journal={IEEE Transactions on Vehicular Technology}, 
    year={2023},
    volume={72},
    number={4},
    pages={5094-5109}}

@inproceedings{Kountouris18,
    author={Kountouris, Marios and Pappas, Nikolaos and Avranas, Apostolos},
    title={{QoS Provisioning in Large Wireless Networks}}, 
    booktitle={Proceedings of the 16th International Symposium on Modeling and Optimization in Mobile, Ad Hoc, and Wireless Networks (WiOpt)}, 
    year={2018},
    pages={1-6}}

@techreport{ETSI_38.212,
    type={{Technical Specification (TS) 138.212 v18.7.0}},
    author={{ETSI}},
    title={{5G; NR; Multiplexing and Channel Coding}},
    year={2025},
    month={July},
    institution={European Telecommunications Standards Institute (ETSI)}}

@techreport{ETSI_38.321,
    type={{Technical Specification (TS) 138.321 v18.6.0}},
    author={{ETSI}},
    title={{5G; NR; Medium Access Control (MAC) Protocol Specification}},
    year={2025},
    month={July},
    institution={European Telecommunications Standards Institute (ETSI)}}

@inproceedings{Wan24a,
    author={Wan, Haoran and Cao, Xuyang and Marder, Alexander and Jamieson, Kyle},
    title={{NR-Scope: A Practical 5G Standalone Telemetry Tool}},
    booktitle={Proceedings of the 20th International Conference on Emerging Networking EXperiments and Technologies (CoNEXT)},
    pages={73–80},
    location={Los Angeles, CA, USA},
    year={2024}}

@manual{5gLENA,
    title={{NR Module User Manual}},
    author={{OpenSim CTTC/CERCA}},
    year={2024},
    note={Release 3.3},
    url={https://cttc-lena.gitlab.io/nr/nrmodule.pdf}}

@article{Irazabal24,
    author={Irazabal, Mikel and Nikaein, Navid},
    journal={IEEE Journal on Selected Areas in Communications}, 
    title={{TC-RAN: A Programmable Traffic Control Service Model for 5G/6G SD-RAN}}, 
    year={2024},
    volume={42},
    number={2},
    pages={406-419}}

@inproceedings{Brunello21,
    author={Brunello, Davide and Johansson S, Ingemar and Ozger, Mustafa and Cavdar, Cicek},
    title={{Low Latency Low Loss Scalable Throughput in 5G Networks}}, 
    year={2021},
    booktitle={Proceedings of the 93rd IEEE Vehicular Technology Conference (VTC2021-Spring)},
    pages={1-7}}

@techreport{ETSI_18,
    type={{White Paper}},
    author={{ETSI}},
    title={{MEC in 5G Networks}},
    year={2018},
    month={June},
    institution={European Telecommunications Standards Institute (ETSI)}}

@inproceedings{Hassani20,
    author={Hassani, Hamid and Gringoli, Francesco and Leith, Douglas J.},
    booktitle={Proceedings of the Mediterranean Communication and Computer Networking Conference (MedComNet)}, 
    title={{Regulating Aggregation Level for Low Latency in 802.11ac}}, 
    year={2020},
    pages={1-7}}

@inproceedings{Zaki15,
    author={Zaki, Yasir and P\"{o}tsch, Thomas and Chen, Jay and Subramanian, Lakshminarayanan and G\"{o}rg, Carmelita},
    title={{Adaptive Congestion Control for Unpredictable Cellular Networks}},
    year={2015},
    booktitle={Proceedings of the ACM SIGCOMM Conference},
    pages={509–522},
    location={London, United Kingdom}}

@inproceedings{Karnam25,
    author={Karnam, Anudeep and John, Jobish and Joshi, Kishor C. and Exarchakos, George and Heemstra de Groot, Sonia and Niemegeers, Ignas},
    booktitle={Proceedings of the IEEE Wireless Communications and Networking Conference (WCNC)},
    title={{Reliability Modeling for Beyond-5G Mission Critical Networks Using Effective Capacity}},
    pages={1-6},
    year={2025}}

@online{L4SNokia,
    title={{L4S}},
    url={https://www.nokia.com/bell-labs/research/l4s/},
    year={2025},
    note={Accessed: May 10, 2026}}

\end{document}